\documentstyle[twocolumn,aps,prl]{revtex}
\input{psfig.sty}
\begin{document}
\draft

\author{Rick Tully and George Reiter}
\address{Physics Department and Texas Center for Superconductivity,University of Houston, Houston, Tx 77204-5506}
\date{\today}
\title{Breakdown of the Universality Hypothesis in Directed Abelian Sandpile Models}
\maketitle
\begin{abstract}
We show that in a broad class of directed abelian sandpile models that had been expected to have the same exponents as the Dhar-Ramaswamy model, 
the avalanche exponent depends upon the details of the interaction, 
calling into question the general existence of universality classes in
self organized critical models.  

\end{abstract}
\pacs{64.60.Lx,05.70.Jk}

	The existence of universality classes in systems exhibiting self-organized critical
behaviour(SOC) is expected  on the basis of the similarity of the SOC state  to the critical state of
equilibrium systems. The particular physical constraints that might determine these classes 
are, however, very much an open question, but they are thought to resemble the universality classes of equilibrium systems, at least as far as symmettry and dimensionality are concerned.   In one dimension, it was shown recently by Paczuski and Boettcher \cite{maya}
that a model for rice piles and another for interface depinning were in the 
same universality class, and they conjectured that the Burridge-Knopoff model \cite{burr}
was contained there as well. Subsequently, Cule and Hwa\cite{cule} described a
related model that was in the same universality class.   
A classification scheme for 2-dimensional abelian sandpile models was suggested 
by Ben-Hur and Biham\cite{ben}. In their paper, models were classified according 
to the "sand current" ,$\vec{J}$, which resulted when an unstable site was 
toppled. Models with
$\vec{J}=0$ and
$\vec{J}\not=0$ were said to be in the "nondirected" and "directed" universality
classes respectively, and a distinction was made between those that were nondirected only on average. Dhar and Ramaswamy\cite{dhar}(DR) had earlier presented an exact solution for a   directed  model in two dimensions, and had 
shown numerically that a closely related directed model had the same exponents. Ben-hur and Biham described having run a third model with the same exponents.
However, in their simulations, only models with nearest
neighbor interactions were considered. Our investigations of longer (but still finite) range
interactions in directed models show a continuous variation of exponent values ranging
 from those of the exactly solvable nearest neighbor model of DR
to that of an exactly solvable infinite range interaction model that we introduce here.

No simple classification is apparent in these models, with exponents
 depending on the details of the interaction, as well as the range.
While it is not unexpected that universality classes in SOC would be
 more complex than those in equilibrium models, this finding calls into question the
 idea of using universality ideas to describe these models at all, at least in two   dimensions. 

 Since the variation in exponents is not large(2.33-2.5), we introduce a simulation method 
 which exploits the properties
of a subclass of the directed models to greatly reduce the deviation from power law
 behaviour
seen in the simulations as a result of  finite system sizes. This allows more 
accurate 
determination
of scaling exponents from relatively small system sizes. We emphasize, 
however, that the standard methods for obtaining the exponents yield the same results.

	To fix notation, we define a sandpile model  by three variable aspects. 
	\begin{description}
	\item[lattice:] In
two dimensions the 
model dynamics occur on a $MxN$ lattice $\mathcal{L}_{MN}$ composed of $d=MN$ 
sites where sand can build up.
This lattice is embedded in an infinite lattice of sites which absorb all sand
which fall on them.
The state of the system at any given time is a
$d$ dimensional integer valued vector whose components represent the height of
individual sites. \item[critical height:] The critical height $h_{c}$ is the maximum 
amount of sand which can sit on a site and not topple. If a site has more than
$h_{c}$
grains it is said to be unstable and will topple.\item[toppling rule:] The toppling (or
relaxation) rule determines how sand is removed from an unstable site $i$ and
redistributed to sites in some predetermined neighborhood ,${\mathcal{N}}_{i}$, of
$i$.
	\end{description}
 One then defines the set of toppling vectors,
\{$\vec{T_{j}}$\} ,where the subscript corresponds to one of the $d$ lattice sites.
The unstable site $i$ with $h_{c}+1$ or more grains is toppled by subtracting
the vector $\vec{T}_{i}$ from the state vector.
The $s^{th}$ component of $\vec{T_{i}}$ is defined as 
\begin{equation}
	[\vec{T}_{i}]_{s} =\left\{\begin{array}{ll}
		h_{c}+1 &\mbox{if  $s=i$}\\
		-f_{i}(s)   &\mbox{$s\not=i$}
		
			\end{array}
	\right. 					
\end{equation}
where
\[
f_{i}(s)=\left\{\begin{array}{ll}a\ positive\ integer &\mbox{if $s\in{\mathcal{N}}_{i}$}\\
	0		 &\mbox{otherwise}
		\end{array}
		\right.
\].
	
For two sites $i$ and $j$, ${\mathcal{N}}_{j}$ is simply a translated
copy of ${\mathcal{N}}_{i}$.
 That is,
all unstable sites topple identically in the sense that
their contents are distributed in the same pattern relative to the
toppling sites. 
The models considered here are further constrained 
so that $\sum_{s=1}^{d}f_{i}(s)\leq h_{c}+1$ where the equality holds when
 ${\mathcal{N}}_{i}\in\mathcal{L}_{MN}$. 
\begin{figure}[h]
\psfig{figure=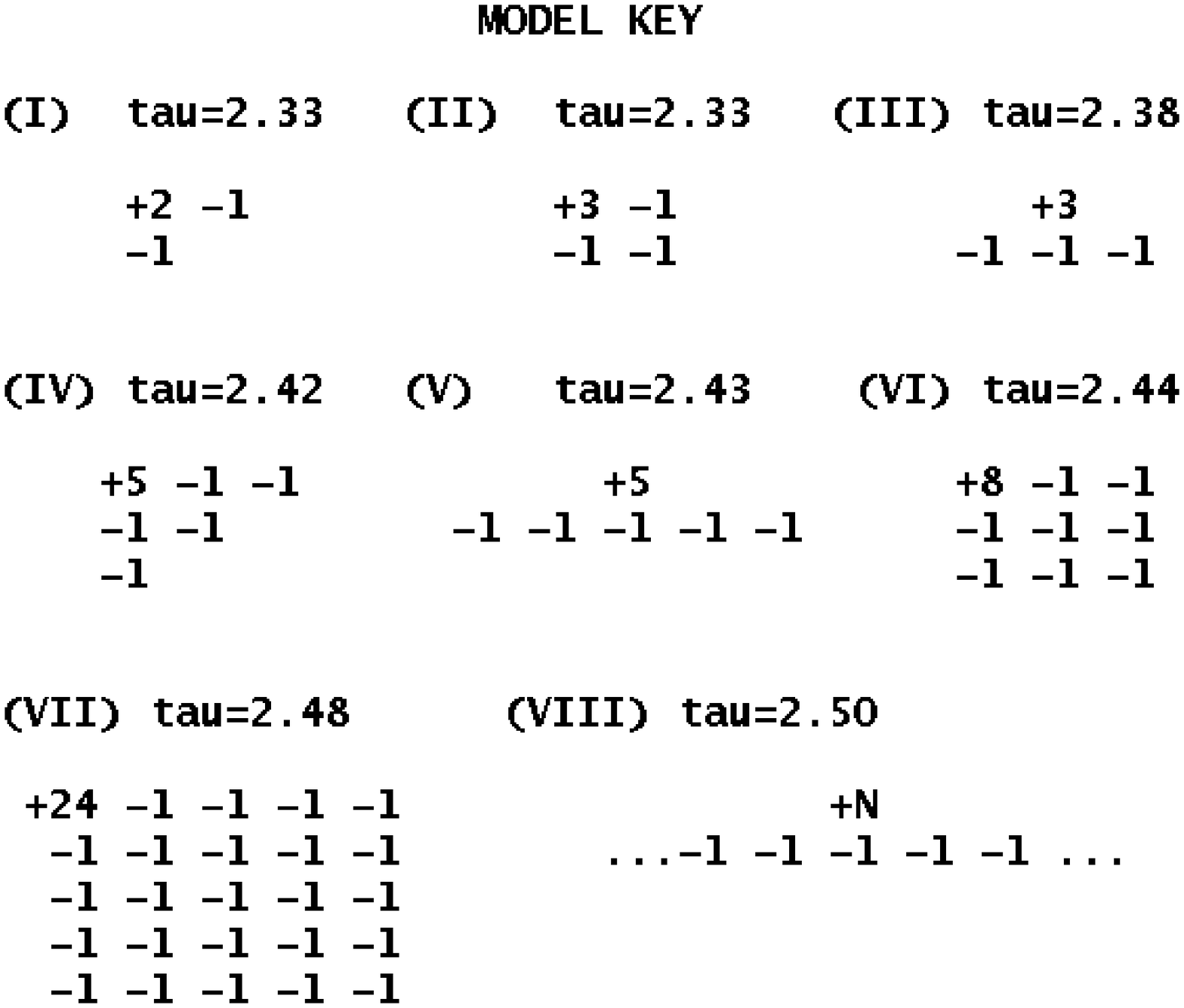,width=80mm,height=50mm}
\caption{Toppling patterns and avalanche exponents for the models studied. The exponents are accurate to at least $\pm$ .01.} 
\label{key}
\end{figure}
In our simulations we considered a variety of directed models with differing
toppling vectors some of which are described in Fig.1. Model I was
previously solved exactly by Dhar and Ramaswamy(DR)\cite{dhar}. 

Abelian sandpile models(ASM) are conventionally driven by adding a grain of sand
 to a random site in a stable configuration. If this site
 becomes unstable then the appropriate toppling vector is subtracted from the
 new state perhaps inducing other sites to become unstable. 
 On the next time step the lattice is reexamined for
 unstable sites which are then toppled. The number of time steps before no new
 unstable sites are found is said to be the avalanche's lifetime. The total
 number of toppled sites is called the avalanche's size or mass. 
 Once the system relaxes, another
grain is randomly added to the new stable state.
DR showed that once an ASM 
is driven into the stationary state by this method, all reccurent 
configurations occur with equal probability. So in theory at least, one could
get the same statistics by adding sand to random 
members of the recurrent set. Typically this is
no small task since for the general ASM the recurrent set is remarkably
complex. However, there is an infinite set of relaxation rules falling under
the directed model classification which posess trivial recurrent sets.
It can be shown that if there exists an orthogonal transformation which leaves
the toppling matrix, $\Delta_{ij}=[\vec{T}_{j}]_{i}$, triangular, 
the recurrent set
will consist of all possible stable configurations. \cite{thesis,dhar2}. 
All of the models we simulated posess this trivial stationary state.

\begin{figure}[h]
\psfig{figure=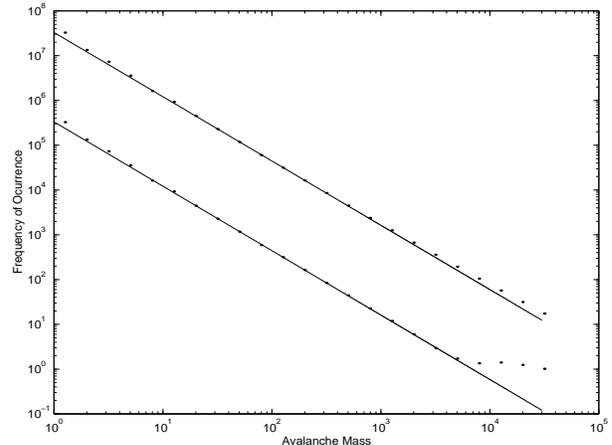,width=80mm,height=60mm}
\caption{The lower curve is calculated for model V by seeding at a fixed site and using a random background on a 500x1000 lattice, the upper curve by seeding at random locations on a 3000x7000 lattice.  The exponent in this case is 2.43 }
\label{face}
\end{figure}

 Since an avalanche may be seeded
at a site near a boundary, even small avalanches may be effected by the finite 
system size. However, given the nature of these directed model's stationary
states and the fact that all sites in the lattice are equivalent 
to the extent that the system is infinite, one may reduce
edge effects by introducing sand at a fixed site a maximum distance
from the boundaries in randomly constructed stable states. Mass data 
from the simulations was obtained by seeding avalanches  at the top
left lattice site for models I, II, IV, VI, and VII and in the middle of the top
row for for models III, V, and VIII. This mass was histogrammed,
logarithmically binned and fitted to distributions of the form
$\mathcal{P}$$(s)$$\propto$ $s^{1-\tau}$ where $\mathcal{P}$$(s)$ is the probability of
the occurence of an avalanche with $s$ toppled sites.
As can be seen from the plots in Fig.\ref{face}, in which model V was used,  the deviation from power 
law behavior in the simulation where sand is dropped on a fixed 
site is much more marked than in the simulation with a random 
drop site. This eliminates much of the guess work in finding a 
suitable endpoint for the linefitting routine. Notice also that the
region of power law behavior in the run with the fixed drop site is at least
as large as that of the run with the random drops even though it has a
considerably smaller lattice. 
  
\begin{figure}[h]
\psfig{figure=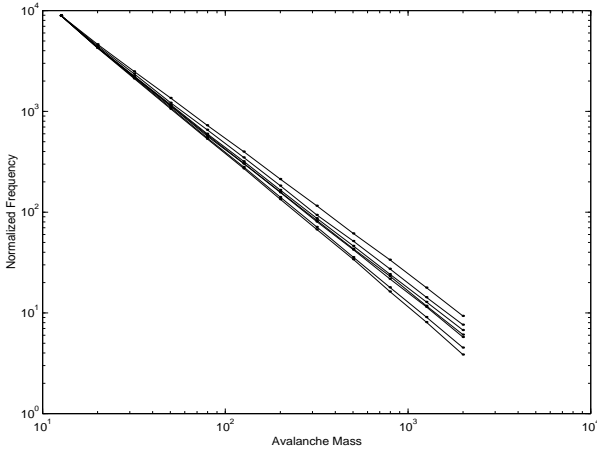,width=80mm,height=60mm}
\caption{Data for the models described in Fig 1. The smallest exponent is from Models I and II, the largest from a simulation of an approximation to the infinite range model described in the text with an N of 100 and $\nu$ of 2. The data has been normalized to make the frequency in the first bin the same for all datasets.}
\label{comp}
\end{figure}

 For the different models we found a range of
values for $\tau$ from 2.33 to 2.50. The data is summarized in Fig.\ref{comp}, and  exponents are given in Fig. \ref{key}.  Although a general trend of 
increasing $\tau$ with increasing $h_{c}$ was observed,
 it was also found that $\tau$ varied with changes in ${\mathcal{N}}_{i}$
 for a fixed $h_{c}$.

\begin{figure}[h]
\psfig{figure=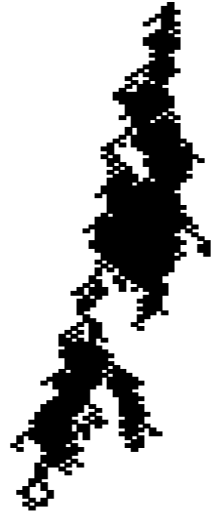,width=80mm,height=70mm}
\caption{This is an avalanche for model III chosen to illustrate the hole formation.  Note the
presence of untoppled sites(white) within the boundaries of the avalanche}
\label{hole}
\end{figure}

 In model II, avalanches occur with the same distribution as the exactly
 solvable model I. This is understandable by observing that
 neither model allows untoppled sites, or "holes", in the interiors of their
 avalanche's constant time surfaces. It was this fact which was essential 
 to DR's
 mapping of the evolution of the perimeter of an avalanche onto the problem
 of annhilating random walkers. 
  Model II,  may be similarly mapped, although
 with a different distribution over the walker's step sizes. The difference 
 in possible step sizes affects the variance of the walk, but not the exponent
  for the time of intersection.  
 The distributions for the lifespans of these walkers
 are then the same as that of the avalanches and are identical for both
 models. However, it is possible for another choice of ${\mathcal{N}}_{i}$ with
  $h_{c}=3$ to produce avalanches with holes and complicated perimeters  
 not suitable for description by random walkers. Model III is one such 
 case  as can be seen from  Fig.\ref{hole}, and indeed,  a different value for $\tau$(see Fig.\ref{key}.) 
 is obtained from its simulation. In general, as $h_c$ increases and the
  toppling vectors change, different patterns of holes are allowed, and different exponents result.  This suggests that any classification
 of directed models must include a description of hole formation.

 An analytically 
 tractable model has been devised to explore
 the limiting case of infinite $h_{c}$.  The model lives on a $2d$ $MxN$ lattice with $h_{c}=N+\nu-1$ where $\nu$
 is some positive integer. A site becomes unstable if it possesses $N+\nu$
 or more grains. The $i_{th}$ unstable site is relaxed by subtracting the
 toppling vector $\vec{T}_{i}$. The components of $\vec{T}_{i}$ are given
 below.
 \begin{equation}
	[\vec{T}_{i}]_{s} =\left\{\begin{array}{ll}
		N+\nu &\mbox{if  $s=i$}\\
		-1   &\mbox{for $s\in$ the next row}\\
		0		&\mbox{$s\not=i$ and $s\not\in$ the next row}
		
			\end{array}
	\right. 					
\end{equation}
In words, if a site becomes unstable and topples, $N+\nu$ grains of its sand
are removed and one grain is added to every site in the next row so that
$\nu$ grains are lost with each topple. Labeling the lattice sites by the
convention which calls the upper left lattice site one and numbers sites across
the first row to $N$ beginning the next row with $N+1$ and so on, one finds the
toppling matrix to be lower triangular. As stated earlier this implies that the critical
state consists of all stable configurations placing this model in the same
category as the rest of the models considered in this treatment.

Note that with this model and for an avalanche beginning with a single unstable 
site, the constant time surfaces and therefore toppling activity
are restricted to a single row. This allows
the row by row evolution of an avalanches' mass to be described by the properties 
of a stationary Markov chain with the row number serving also as the time
parameter. 

Let ${\mathcal{S}}_{n}$ be
the random variable associated with the number of sites which topple on the
$n^{th}$ row.Since a sites height in the stationary state can take values from
$0$ to $N+\nu-1$ with equal probability, the conditional probability that
 a single stable site has $z$
grains and topples given that $i$ topple on the previous row is 
\begin{equation}
{\mathcal{P}}[{\mathcal{S}}_{n+1}=$1$|{\mathcal{S}}_{n}=i]=\mathcal{P}[$z$\geq
N+\nu-$i$]
=\frac{i}{N+\nu}
\end{equation}.

The probabilities of different sites toppling on the same row are independent so
the transition probabilty, $p_{ij}$, of $j$ sites toppling given that $i$ toppled on the
previous row can be written as,
\begin{equation}
{\mathcal{P}}[{\mathcal{S}}_{n+1}=j|{\mathcal{S}}_{n}=i]=
\left(\begin{array}{c}N\\j\end{array}\right) 
{\left(\frac{i}{N+\nu}\right)}^{j}
{\left(\frac{N+\nu-i}{N+\nu}\right)}^{N-j}
\end{equation}
It is well known that a binomial distribution of this form approaches a Poisson
distribution in the limit of large $N$ so that the above transition
probability may be reexpressed as,
\begin{equation}
\lim_{N\rightarrow \infty}{p}_{ij}=\frac{i^{j}}{j!} e^{-i}
\end{equation}
.

Letting $\mathcal{T}$ be the random variable associated with the avalanche 
lifetime, (or equivalently the
number of rows which have at least one toppling event), it is clear that
\begin{equation}
{P}[{\mathcal{T}}\leq n]={P}[{\mathcal{S}}_{n+1}=0]
\end{equation}
and so
\begin{equation}
{P}[{\mathcal{T}}\leq n]={P}[{\mathcal{S}}_{1}=0]+
{P}[{\mathcal{S}}_{n+1}=0|{\mathcal{S}}_{1}=1]{P}[{\mathcal{S}}_{1}=1]
\end{equation}
for avalanches begun by adding a single grain to the first row.
${P}[{\mathcal{S}}_{1}=0]$ and ${P}[{\mathcal{S}}_{1}=1]$ are constants
for a fixed system size and ${P}[{\mathcal{S}}_{n+1}=0|{\mathcal{S}}_{1}=1]$ 
can be expanded
in terms of the transition probabilities as,
\begin{equation}
{P}[{\mathcal{S}}_{n+1}=0|{\mathcal{S}}_{1}=1]=
\sum_{s_{n}=0}^{N} \sum_{s_{n-1}=0}^{N} \ldots
\sum_{s_{2}=0}^{N} p_{s_{n}0} p_{s_{n-1}s_{n}} \ldots p_{1s_{2}}
\end{equation}.  
Carrying out the above sums for
different $n$'s leads to the following sequence of conditional probabilities.
\begin{equation}
\begin{array}{l}
{P}[{\mathcal{S}}_{1}=0|{\mathcal{S}}_{1}=1]=0 \\
{P}[{\mathcal{S}}_{2}=0|{\mathcal{S}}_{1}=1]=e^{-1} \\
{P}[{\mathcal{S}}_{3}=0|{\mathcal{S}}_{1}=1]=e^{{P}[{\mathcal{S}}_{2}=0|{\mathcal{S}}_{1}=1]-1}\\
\vdots \\
{P}[{\mathcal{S}}_{n+1}=0|{\mathcal{S}}_{1}=1]=e^{{P}[{\mathcal{S}}_{n}=0|{\mathcal{S}}_{1}=1]-1}
\end{array}
\end{equation}

Expanding  ${P}[{\mathcal{S}}_{n}=0|{\mathcal{S}}_{1}=1]$ in powers of 
$\frac{1}{n}$ and using
the above recursion to find the coefficients, one finds that to first order,
\begin{equation}
{P}[{\mathcal{S}}_{n}=0|{\mathcal{S}}_{1}=1]=1-\frac{2}{n}
\end{equation}
. So that for long
lived avalanches in large systems, we find that the probability
of an avalanche lasting longer than $n$ time steps is 
\begin{equation}
{P}[{\mathcal{T}}> n]\simeq \frac{A}{n^{\alpha}}
\end{equation}
where $A=2{P}[{\mathcal{S}}_{1}=1]$ and $\alpha=1$. 
The average flux of grains in and out of a row in the
steady state must be constant. This implies that the probability
of the mass of an avalanche being greater than $m$ is proportional
to $m^{2-\tau}$ where $\tau$ is given by the relationship
$\tau=2+\alpha/(1+\alpha)$ \cite{dhar}. This value of $2.5$ for $\tau$ matches
the value obtained by numerical simulation to within $\pm .01$ for $N=100$ and $\nu$=2.

From the point of view of the analogy with equilibrium critical systems, 
the continuous variation of the avalanche exponent with the range and details 
of the interaction is not understandable. There are of course equilibrium models with varying exponents, 
 such as the Ashkin-Teller model, but we see no connection of the SOC models we have considered with these models. 
 Varying exponents have also been obtained in dissipative SOC models, such as the one introduced by 
  Olami, Feder, and Christensen\cite{olami}, but in these models, the dissipation, which effectively 
  links every site to the boundary, is a long ranged interaction, and so the variation of exponents 
  is at least not in contradiction with the intuition gained from equilibrium systems.\cite{tang}

However, in our case, the models intermediate between the DR model and the mean field
model we have introduced are all short ranged in the usual sense of equilibrium critical phenomena. 
They should all have the same critical exponents, which is  Ben-Hur and Biham's
 conjecture, and they do not. We do not think that the natural point of view from this perspective, 
 that we are just not yet in the critical region for the data we have shown, and they will all 
 eventually cross over to a common exponent of 2.33,  
 is tenable. By using the fact that the recurrent set is the stable cube for our models, we eliminate the 
 problem of long transients or uncertainty in the exponent due to edge effects. For model three, the system 
 size used is about 100 times the interaction range, which is the only scale in the problem. 
 If there was going to be a cross-over, we should have seen it. We conclude that there 
 is no universality in those 2-d directed models  whose avalanches from 
 recurrent states contain  holes. The extent to which the breakdown of universality extends to other 
 SOC models remains to be seen\cite{bih}.   
\centerline{\bf Acknowledgement}
We thank Maya Paczuski for her encouragement and many useful conversations. This work was supported by the Texas Center for Superconductivity through a grant from the state of Texas.


%

%
%


\begin{references}
\bibitem{maya} Stefan Boettcher and Maya Paczuski, Phys. Rev. Letts. {\bf 77},111 (1996)  
\bibitem{burr} R. Burridge and L. Knopoff, Bull. Seismol. Soc. Am {\bf 57}, 341 (1967)
\bibitem{cule} Dinko Cule and Terence Hwa, Phys. Rev. B {\bf 57}, 8235 1998
\bibitem{ben} A. Ben-Hur and O. Biham, Phys. Rev. E, {\bf 53}, R1317 (1996)
\bibitem{dhar} Deepak Dhar and Ramakrishna Ramaswamy, Phys. Rev. Letts. {\bf 63}, 1659 (1989)
\bibitem{dhar2} D. Dhar, Phys. Rev. Letts. {\bf 64}, 1613 (1990)
\bibitem{thesis} thesis, Rick Tully, (to be published)
\bibitem{olami} Z. Olami, H.J.S. Feder and K. Christensen, Phys. Rev. Letts {\bf 68}, 1244, (1992);K. Christensen and Z. Olami, Phs. Rev. A {\bf 46}, 1829 (1992)
\bibitem{tang} An analysis  
 of the reasons for the variation of the exponent in this case is given
  in A.Alan Middleton and Chao Tang, Phys. Rev. Letts. {\bf 74}, 742 (1995)
\bibitem{bih} A breakdown of universality in a non-abelian model has been reported in
O.Biham,E.Milshtein and S. Solomon, cond-mat/9805206.  
\end{references}
\end{document}